\documentclass[useAMS,usenatbib]{mn2e}
\usepackage{amsmath,fleqn,graphicx,txfonts}
\usepackage{times}
\usepackage{epsf}
\usepackage[dvips]{epsfig}
\usepackage{footnote}
\usepackage{threeparttable}
\usepackage{array}
\usepackage{anysize} 
\usepackage{natbib}
\label{firstpage}

\title[Origin and Chemical Evolution of NGC3201]{The Origin and Chemical Evolution of the Exotic Globular Cluster NGC3201}

\author[C. Mu\~{n}oz et al. 2013]{
 C. Mu\~{n}oz \thanks{cesarmunoz@astro-udec.cl}, D. Geisler
 and S. Villanova\\
   Departamento de Astronom\'{i}a, Casilla 160-C, Universidad de
  Concepci\'{o}n, Concepci\'{o}n, Chile \\
}
\begin{document}

%-----------------------------------------------------------------------------------------------------------------------------------------------------------------------
%--------------------------------------------------------------------------- ABSTRACT ------------------------------------------------------------------------------------
%-----------------------------------------------------------------------------------------------------------------------------------------------------------------------

\pagerange{\pageref{firstpage}--\pageref{lastpage}} \pubyear{2013}

\maketitle

\begin{abstract}
NGC3201 is  a globular cluster (GC)  which shows  very peculiar kinematic characteristics including an extreme radial velocity and  a highly retrograde orbit, strongly suggesting an extraGalactic origin.\\
 Our aims are to  study NGC3201 in the context  of multiple populations (MPs), hoping to constrain
possible candidates for the self-enrichment by studying  the chemical abundance pattern,
as well as adding insight into the origin of this intriguing cluster.\\
We present a detailed chemical abundance  analysis of eight red giant branch (RGB) stars using high resolution spectroscopy.
We measured 29 elements and   found [Fe/H]=-1.53$\pm$0.01, we cannot rule out a metallicity spread of $\sim$0.12 dex,  and an $\alpha$-enhancement typical of halo GCs. However significant spreads are observed in the abundances of all light  elements except for Mg.
We confirm the presence of an extended Na-O anticorrelation. n-capture elements generally are dominated by the
 r-process, in good agreement with the bulk of Galactic GCs. The total (C+N+O) abundance is slightly
 supersolar and requires a small downward correction to the isochrone age, yielding 11.4 Gyr.\\
 Kinematically, NGC3201 appears likely to have had an extraGalactic origin but its chemical evolution is similar to  most other, presumably native, Galactic GCs.
\end{abstract}

%-----------------------------------------------------------------------------------------------------------------------------------------------------------------------
%-----------------------------------------------------------------------------------------------------------------------------------------------------------------------
%-----------------------------------------------------------------------------------------------------------------------------------------------------------------------
%-----------------------------------------------------------------------------------------------------------------------------------------------------------------------
%--------------------------------------------------------------------INTRODUCTION----------------------------------------------------------------------------
%-----------------------------------------------------------------------------------------------------------------------------------------------------------------------

\section{Introduction}
\label{sec:intro}
GCs  are of paramount importance  in the investigation of a wide variety of fundamental astrophysical phenomena.  They are the perfect laboratories for studying a wide variety of fundamental problems in stellar and galactic astrophysics. Our knowledge of stellar evolution is grounded in the detailed comparison of GC color-magnitude diagrams with model isochrones.  They   are the oldest  objects known whose ages are relatively easily and accurately measured, setting a firm lower limit to the age of the Universe.

Only a few years ago  GCs, with the lone exception of $\omega$ Cen, were considered Simple Stellar Populations, with all stars in a cluster believed to share the same age and che\-mical composition. This simplicity
has been shattered by the discovery of increasing complexity in the latter parameter for a growing number of GCs, spawning the new and excit\-ing field of multiple populations (MPs). Both photometric and spectroscopic 
techniques are employed to trace and understand this phenomenon.
The   development  of  larger tele\-scopes, more 
 sensitive detectors and high resolution multiobject spectrographs has allowed a large sample of giants in each of
a growing number of  GCs to be studied with high resolution spectroscopy, leading to an ever-increasing database of elemental abundances. The emerging picture, although quite complicated in detail, does 
show several salient features.

In particular, 
light elements (C, N, O, Na, Mg, Al)  are associated with the  MP phenomenon.
The most typical and best-studied characteristic  is the Na-O anticorrelation. Indeed, the ubiquity of this anticorrelation in their sample of 19 GCs studied prompted \citet{carretta2009a,carretta2009b}   to introduce a new, chemical definition of a GC as those objects which exhibit such a feature. 
This spread in the light elements  must be due to self-enrichment  that happens within a  GC in the  early stages of its formation, when a second gene\-ration of stars was born from gas polluted by ejecta of evolved stars 
of the first generation \citep{caloi}.
Several kinds of polluters for the light elements have been proposed: intermediate mass AGB stars \citep{dantona}, fast 
rotating massive MS stars \citep{decressin}  and massive binaries \citep {mink}. In addi\-tion, a few of the most massive clusters such as $\omega$ Cen \citep{johnson2008,marino2011a}, M54 \citep{carretta2010b} and  M22 \citep{marino2011b}  are now known to show significant spreads in Fe as well. Indeed, the existence of such  a metallicity spread has also been claimed in NGC3201 \citep{gonzalez,simmerer}, but remains controversial. Such clusters must also have been able to retain SNeII and/or SNeIa ejecta \citep{marcolini}, as well as material ejected from polluters at lower velocity which led to the
Na-O anticorrelation.
Ejecta from evolved stars (cycled through a temperature of T\textgreater 10$^{7}$ K where hot H-burning occurs)  is returned to the 
 interstellar  medium of the GC and mixes with the remaining  primordial gas left over from the first star formation epoch. Given a sufficiently deep potential well to retain this ejecta,   which is the critical parameter, a  second  generation of star formation can occur once the appropriate conditions arise \citep[e.g.][]{cottrell,carretta2010c}.

NGC3201 is unique  among the Galactic GCs kinematically. It has the  most extreme radial velocity (495 $km$ $s^{-1}$).
 \citet{gonzalez}  calculated an orbital velocity of 250 $km$ $s^{-1}$ around the Galactic center, but in a retrograde sense. This has been taken as strong evidence of a possible extraGalactic formation with subsequent  capture by the MW  \citep {rodgers,van}.
 
Examining   the color-magnitude diagram (CMD) of  NGC3201  \citep[e.g.][]{layden}  clearly shows that it has a very populated and extended horizontal branch (HB). 
 According to \citet{dantona}, the extension 
of the HB in a GC is proportional to the amount of helium variation due to self-pollution among its stars. Therefore, the large extension of the HB in NGC3201   suggests that it should display a large spread in the light elements.

 \citet{kravtsov} and \citet{carretta2010d}  found radial inhomogeneities in the stellar populations of NGC3201, Kravtsov found radial variations in the CMD, that could not be explained by reddening. \citet{carretta2010d} compared their spectroscopic sample of 100 stars with Na and O abundances with photometric data and found that the giant stars of the second generation have a tendency to be more concentrated than stars of the first generation, as expected in the self-pollution scenario.

Several chemical analyses have been carried out on NGC 3201 stars using high-resolution spectroscopy. Unfortunately, these are not sufficient to give a definitive picture of the chemical evolution of this interesting  GC. \citet{gonzalez}  measured  the abundances of several  chemical elements in eighteen stars  but  with large errors, based on CTIO 4m data.  \citet{carretta2009a}   observed a large sample of
 some 100 stars but only measured  the abundances of light elements (O , Na, Mg, Al and Si), as well as Fe. \citet{simmerer}  observed a sample of 24 stars   but only measured  Fe. They  found an  intrinsic spread in iron. We discuss this point in detail in  section 4.1. Our sample (only eight stars), although much smaller than either   the Gonzalez, Carretta or Simmerer  samples,  was observed  with high resolution as well as excellent signal to noise (see section 2), which allows us to measure the abundances of more elements (29 total) with much smaller errors than in the previous studies.\\

In  Section 2 we describe the data reduction and  in Section 3 the methodology we used to obtain the chemical abundances. 
In Section 4 we present our results including iron-peak elements, alpha elements, the Na-O anti-correlation, the Mg-Al anti-correlation,  the total (C+N+O) abundance, the correction to age estimates due to the former,
and heavy elements. In Section 5 we discuss the origin of  NGC3201. Finally in Section 6 we  give a summary of our main fin\-dings.\\

\begin{figure}
%\centering
  \includegraphics[width=3.2in,height=3.2in]{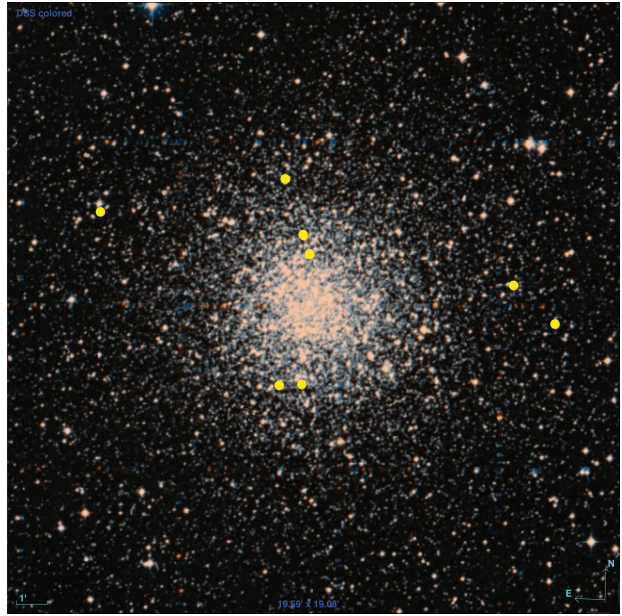}
  \caption{ The spatial distribution of the stars observed (yellow filled circles)  in NGC3201. (picture from SDSS).}
\label {f1}
\end{figure}
 \vspace*{-1cm}

%-----------------------------------------------------------------------------------------------------------------------------------------------------------------------
%------------------------------------------------------------------------OBSERVATIONS AND DATA REDUCTION--------------------------------------%-----------------------------------------------------------------------------------------------------------------------------------------------------------------------

\section{Observations and data reduction}
\label{sec:data}

Our dataset consists of high resolution spectra obtained with the MIKE spectrograph mounted on the Magellan  Clay tele\-scope at Las Campanas \hspace{0.1cm} Observatory, collected in 
March 2011. Our targets include  9 stars  between K= 8.6   and K= 9.8 selected from the upper RGB. 
Our selection criteria included an attempt to cover  a wide spatial distribution around  the
 GC (see Figures 1 and 2).
 Each star was observed with the blue and red arms of the spectrograph, and spectra cover  a wide range from 3900-8079 $\AA$  with  a resolving power of $R\sim41,000$
in the blue and $R\sim32,000$ in the red. Exposure times were between 240-600 seconds. For  some giants,  we stacked several spectra in order to increment the S/N. 
The S/N is between 40-90 at 6500 $\AA$.

Data were reduced using a pipeline from LCO for this instrument (http://code.obs.carnegiescience.edu/mike).
Data reduction  includes bias subtraction, flat-field correction, wavelength calibration, sky subtraction, and spectral rectification.

We measured radial velocities using the fxcor package in IRAF and a synthetic spectrum as a template. This spectrum was generated using the mean of the atmospheric parameters of our stars.
One star (0437-0223437 NOMAD-1 Catalog)
 shows a very discrepant radial
velocity (20.76 $km$ $s^{-1}$), and so was rejected as a non-member.  The mean heliocentric radial velocity  value for our remaining 8 targets is 
$495.18 \pm  0.81$  $km$ $s^{-1}$ , while the dispersion is $2.14$ $km$ $s^{-1}$. The mean radial velocity is in excellent agreement 
with the values in the literature: \citet{carretta2009a}  found a  value of  $494.57 \pm  1.07$  $km$ $s^{-1}$ and \citet{harris} quote a value of $494 \pm 0.2$ $km$ $s^{-1}$. 
Table 1 lists the basic parameters  of the selected stars: ID (the NOMAD-1 Catalog), the J2000 coordinates (RA and Dec), V, B magnitudes \citep{cote}, K, J magnitudes
 (2MASS), heliocentric radial velocity,  $T_{eff}$, log(g), micro-turbulent velocity ($v_{t}$) and metallicity ([Fe/H]). 

%\vspace{1 cm}

\begin{figure}
\centering
    \includegraphics[width=3.2in,height=3.2in]{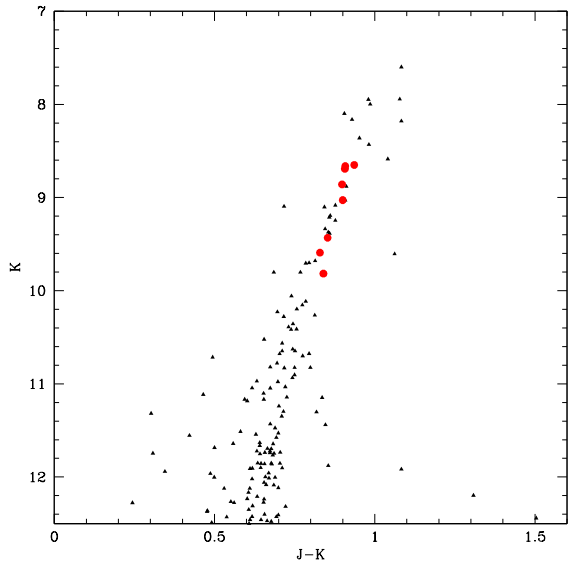}
  \caption{ The CMD of NGC3201 from NOMAD catalogue \citep{zacharias}  with the observed RGB stars indicated as red filled circles.}
  \label{f2}
\end{figure}

%-----------------------------------------------------------------------------------------------------------------------------------------------------------------------
%------------------------------------------------------------------------------------------ABUNDANCE ANALYSIS------------------------------------------
%-----------------------------------------------------------------------------------------------------------------------------------------------------------------------

\section{Abundance analysis}
\label{sec:abundance}
The chemical abundances for  Ca, Ti, Cr, Fe and Ni  were obtained using equivalent widths (EWs) of the spectral lines. See 
\citet{marino2008} for a more detailed explanation of the method we used to measure the EWs. For the other elements 
(C, N, O, Na, Mg, Al, Si, Sc, V, Mn, Co, Cu, Zn, Y, Zr, Ba, La, Ce, Pr, Nd, Sm, Eu, Dy, Pb), whose lines are affected by 
blending, we used the spectrum-synthesis method. We calculated 5 synthetic spectra having different abundances for each line, and estimated the best-fitting value as the one that minimizes the rms scatter. Only lines not contaminated by telluric lines were used. 

Atmospheric parameters were obtained in the follow\-ing way.
First,  $T_{eff}$ was derived from the B-V color \citep{cote} using the relation of \citet{alonso}  and the reddening (E(B-V)=0.24) 
from \citet{harris}. Surface grav\-ities (log(g)) were obtained from the canonical equation:

\vspace{0.3cm}
$log (g/g_{\sun}) = log (M/M_{\sun}) + 4  log (T_{eff} /T_{\sun}) - log (L/L_{\sun})$
\vspace{0.3cm}

\noindent Where the mass  was  assumed to be 0.8 $M_{\sun}$, and the luminosity  was obtained from the absolute 
magnitude $M_{v} $ assuming an apparent distance modulus of  $(m-M)_{V}=14.20$ \citep{harris}. The bolometric correction (BC)
 was derived by adopting the $BC$:$T_{eff}$ relation from \citet{alonso}. Finally, the micro-turbulent velocity $(v_{t})$ was obtained 
 from the relation of  \citet{marino2008}.
These atmospheric parameters were  taken as initial \hspace{ 0.1cm} estimates and were refined during the abundance analysis. As a first step, 
atmospheric models were calculated using ATLAS9 \citep{kurucz} and assuming the initial estimate of $T_{eff}, log(g), v_{t}$, and
 the [Fe/H] value from \citet {harris}.
Then  $T_{eff}, v_{t}$, and $log(g)$ were adjusted and new atmospheric models calculated iteratively in order to remove trends 
in excitation potential  and equivalent width vs. abundance for $T_{eff}$ and $v_{t}$ respectively, and to satisfy  the ionization
 equilibrium for log(g). FeI and FeII were used for this latter purpose. The [Fe/H] value of the model  was changed at each iteration according 
 to the output of the abundance analysis. The local Thermodynamic Equilibrium (LTE) program MOOG \citep{sneden1973} was used for
  the abundance analysis. 
The linelist for the chemical analysis was described  in previous papers \citep[e.g.][]{villanova2011}. The adopted solar abundances we 
used are reported in Table 2.

Abundances for C, N and O were determined all together interactively in order to take into account  molecular coupling
 of these elements. Our targets are objects evolved off the main sequence, so some evolutionary mixing  is expected. This can affect
  the primordial C, N and O abundances separately, but should not affect the total (C+N+O) content because these elements are transformed one into the other
during the CNO cycle. Since all of our  stars are  in the same evolutionary phase,  relative C, N, O 
abundances should also be unaffected.

An internal\hspace{0.1cm} error \hspace{0.1cm}analysis was\hspace{0.1cm} performed by varying $T_{eff}$, log(g), [Fe/H], and $v_{t}$ and redetermining abundances of star \#0436-0222665, assumed
 to be representative of the entire sample. Parameters were varied by $\Delta T_{eff}=+40$ K,  $\Delta$log(g)=+0.14, $\Delta$[Fe/H]=+0.05 dex, and 
 $\Delta v_{t}=+0.03$ $km$ $s^{-1}$, which we estimated as our internal errors. This estimation of the internal  errors for the atmospheric parameters was performed as in Marino et al. (2008). Results are given
  in Table 3, including the error due to the noise of the spectra. This error was obtained as the average value of the errors of the mean given by MOOG for elements measured via EQW; and for elements whose abundance was obtained by spectrum-synthesis, as the error given by the fitting procedure. $ \sigma_{tot}$ is
    the squareroot of the sum of the squares of the individual errors, while $\sigma_{obs}$ is the mean observed dispersion. 
The light elements C, N, O, Na and Al  show a large excess of $\sigma_{obs}$ over $ \sigma_{tot}$,  indicating  inhomogeneity for these 
elements. For all  other elements except  Dy, the agreement is good, indi\-cating no significant intrinsic spread.\\

\begin{figure}
\centering
    \includegraphics[width=3.2in,height=3.2in]{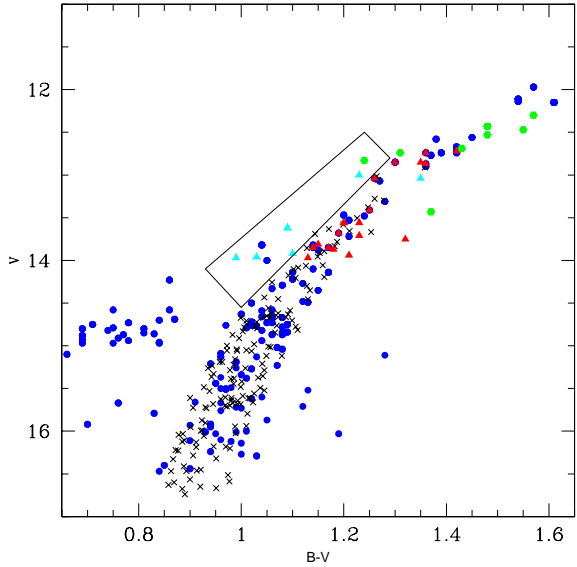}
  \caption{ The CMD of NGC3201, filled green circles are our data, filled blue circles are data from \citet{cote}, black crosses  are data from \citet {carretta2009c}, red filled triangles are data from \citet {simmerer} with [Fe/H]$\geq$-1.56 and cyan filled triangles are data from \citet {simmerer}  with [Fe/H]$\le$-1.59.}
  \label{f2}
 \end{figure}

\begin{figure}
\centering
    \includegraphics[width=3.2in,height=3.2in]{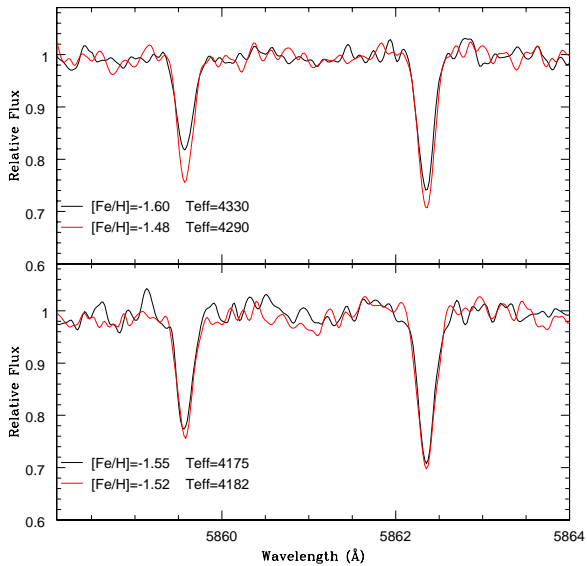}
  \caption{ The upper panel shows the spectral comparison between our most metal-poor  star \#3 and our most  metal-rich star \#4  in two lines of Fe. The bottom panel shows the comparison between stars \#6 y \#7, both with similar [Fe/H] and temperature, for the same Fe lines.}
  \label{f2}
 \end{figure}
%\vspace{0.1cm}

In the case of Sr our measurements give an abundance which is very low,  using the line at 4607.33 $\AA$.  In the literature,  several other lines have been used to measure  Sr, e.g. 4077.71 $\AA$ and 4215.5 $\AA$. However, these lines have  problems due to the large noise present in that part of  our  spectra.
\citet{bergemann}  and \citet{hansen}  found that the 4607.33 $\AA$ line gave widely varying results
when compared with  the line usually used (4077.71 $\AA$) using an LTE model. Because of this, we decided to not use the values of Sr.\\
 The abundance of [Na/Fe]  was corrected by -0.05 dex using  the non-LTE (NLTE) correction (constant for our range of temperature) from  \citet{mashonkina}. For all  other  elements the  selected  lines should  not be  affected significantly by NLTE.

%-----------------------------------------------------------------------------------------------------------------------------------------------------------------------
%------------------------------------------------------------------------------REDDENING----------------------------------------------------------------------------
%-----------------------------------------------------------------------------------------------------------------------------------------------------------------------
\subsection{Reddening}
\label{sec:Red}
The reddening in NGC3201 is a factor to consider given its low galactic latitude (b= $8.6^{\circ}$). The nominal color excess  is E(B-V) = 0.24 \citep{harris}, but a significant differential reddening exists. \citet{gonzalez}  reported a variation in reddening of about 0.1 mag in E$_{B-V}$. \citet{von}   found a differential reddening of 0.2 mag in E$_{V-I}$. \ \

Although, initially, the atmospheric parameters were determined using the nominal reddening, and are thus susceptible to error due to the differential reddening of this cluster, these parameters were refined  interactively, presumably thereby minimizing the effects of reddening on  the measurement of abundances.

%-----------------------------------------------------------------------------------------------------------------------------------------------------------------------
%----------------------------------------------------------------------------------------------------------------------------------------------------------
%-----------------------------------------------------------------------------------------------------------------------------------------------------------------------

\begin{figure*} 
%\centering
  \includegraphics[scale=0.75]{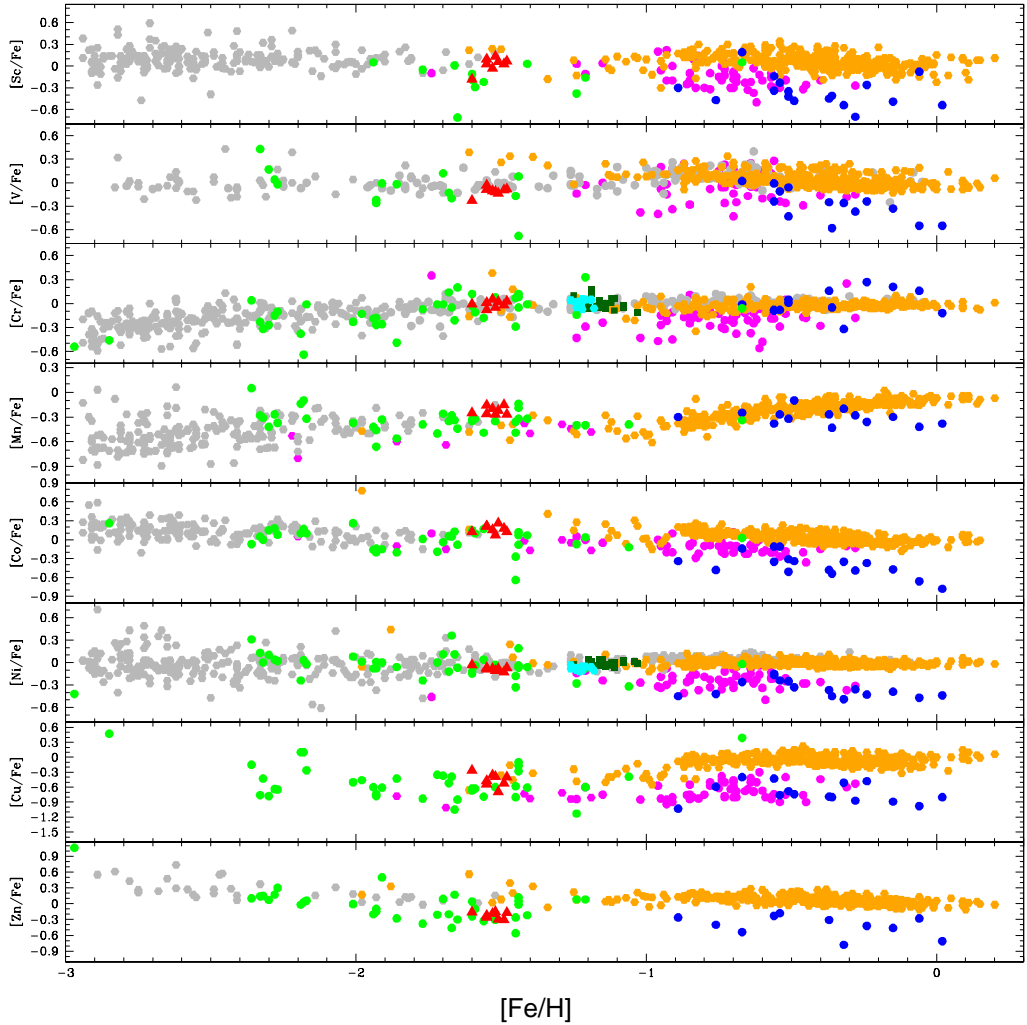}
  \caption{[Sc/Fe], [V/Fe], [Cr/Fe], [Mn/Fe], [Co/Fe], [Ni/Fe], [Cu/Fe], [Zn/Fe] vs [Fe/H]. Filled red triangles are our data  from NGC3201, filled gray circle are halo stars from  \citet{fulbright} and \citet{francois}(for all elements, except for [Cu/Fe]), filled orange circle are  disk stars from \citet{reddy2003,reddy2006}, filled green circle are Draco, Sextans and Ursa minor stars  from \citet{shetrone},  filled magenta circle are LMC stars  (for all elements, except for [Mn/Fe] and [Zn/Fe]) from \citet{pompeia}, filled blue circle are Sagittarius  stars from \citet{sbordone}, filled dark green squares is M4 for [Cr/Fe] and [Ni/Fe]   from \citet{villanova2011}  and filled cyan circle is NGC1851  for [Cr/Fe] and [Ni/Fe] from  \citet{villanova2010}. }
\label {f1}
\end{figure*}

%-----------------------------------------------------------------------------------------------------------------------------------------------------------------------
%-------------------------------------------------------------------------------TABLES----------------------------------------------------------------------------
%-----------------------------------------------------------------------------------------------------------------------------------------------------------------------

\begin{table*}
\caption{Basic parameters of the observed stars.}
\label {t1}
\centering
\begin{tabular}{ l c c c c c c c c c c c c  }

\hline 
\hline
 
$N^{\circ}$ & NOMAD ID& Ra& DEC &B & V & J& K& $RV_{H}$   & $T_{eff}$& log(g)  & $v_{t}$ & [Fe/H] \\

  &&  (h:m:s)&($\,^{\circ}{\rm }$:$^{\prime}$:$^{\prime \prime}$ )& (mag)&(mag)&(mag)&(mag) &(km/s) & (K) &(dex)& (km/s) \\
 \hline
1& 0436-0223885 & 10:18:15.8 & -46:21:51.2 & 14.800 &13.430 & 10.657 & 9.817  &493.57 & 4480 & 1.13 & 1.49 & -1.51 \\ 
2&0435-0223784  &  10:16:51.7 & -46:25:22.6 & 14.070& 12.830 &10.421 & 9.592  &492.85& 4410 & 0.92 & 1.58 &-1.53 \\
3&0435-0223980 & 10:16:59.2 & -46:24:11.1 & 14.050& 12.740 &  10.285 & 9.432 &491.57 & 4330 &0.60 & 1.55 & -1.60\\
4&0436-0222665 & 10:17:38.1 & -46:22:39.2 &  14.120 & 12.690 & 9.928 & 9.028 &495.60 &4290 & 0.94 &1.64  &-1.48\\
5&0435-0225585 & 10:17:42.5 & -46:27:15.4 & 14.010 & 12.530 & 9.757 & 8.859 &497.90&4280 & 0.72 & 1.71 & -1.49\\
6&0435-0225390 & 10:17:38.6& -46:27:16.0   & 14.020 & 12.470 & 9.586 & 8.650 &496.64&4175 & 0.45 & 1.76 & -1.55\\
7&0436-0222791 & 10:17:41.5 & -46:20:51.1& 13.870 & 12.300 & 9.571 & 8.663 &497.53& 4182 & 0.62 & 1.74 & -1.52\\
8&0436-0222622 & 10:17:36.9 & -46:23:12.0 &  13.910 & 12.430 & 9.599 & 8.692 &495.74&4210 & 0.65 & 1.68 &-1.55\\

\hline
\end{tabular}
\end{table*}

\begin{table*}
%\begin{threeparttable}
\caption{Columns 2-9: abundances of the observed stars. Column 10: mean abundance for the cluster. Column 11: abundances adopted for the Sun in this paper. Abundances for the Sun are indicated  as log$\epsilon$(El.).}
\begin{threeparttable}
\label {t2}
\begin{tabular}{ l c  c c  c  c c  c  c c  c }

\hline 
\hline
El. & 1 & 2  & 3 & 4  & 5 & 6 & 7 & 8 & Cluster \tnote{1} &Sun   \\
\hline 
$ [C/Fe] $   &   -0.97  &	-0.76  &	-0.34  &	-0.35 &	-0.40  &	-0.90  &	-0.43  &	-0.62 & -0.60$\pm$0.09   &  8.49\\
$ [N/Fe] $   &   1.09   &       0.98   &        0.31   &	0.29  &	-0.05  &	0.53   &	0.46   &	0.77  & +0.55$\pm$0.13   & 7.95\\
$ [O/Fe] $   &   -0.36  &	-0.16  &	0.15   &	0.40  &	0.27   &	0.37   &	0.25   &	0.11  & +0.13$\pm$0.09   & 8.83\\
$ [CNO/Fe] $ &   0.14   &	0.13   &	0.09   &	0.29  &	0.15   &	0.27   &	0.18   &	0.15  & +0.18$\pm$0.02   & 9.01\\
$ [Na/Fe]_{NLTE} $  &	 0.6   &	0.45  &	-0.35  &	-0.2 &	-0.25  &	-0.05   &	0.05   &	0.28  & +0.07$\pm$0.12   & 6.32\\
$ [Mg/Fe] $  &	 0.28   &	0.25   &	0.54   &	0.41  &	0.38   &	0.34   &	0.44   &	0.37  & +0.38$\pm$0.03   & 7.56\\
$ [Al/Fe] $  &	 0.92   &	0.93   &	0.36   &	0.19  &	0.15   &	0.14   &	0.20   &	0.63  & +0.44$\pm$0.12   & 6.43\\
$ [Si/Fe] $  &	 0.23   &	0.27   &	0.12   &	0.22  &	0.25   &	0.27   &	0.34   &	0.29  & +0.25$\pm$0.02   & 7.61\\
$ [Ca/Fe] $  &	 0.31   &	0.31   &	0.40   &	0.25  &	0.33   &	0.30   &	0.33   &	0.33  & +0.32$\pm$0.01   & 6.39\\
$ [Sc/Fe] $  &	 0.06   &	-0.03  &	-0.18  &	0.07  &	0.03   &	0.03   &	0.13   &	0.09  & +0.03$\pm$0.04   & 3.12\\
$ [Ti/Fe] $\tnote{2}  &	 0.22   &	0.26   &	0.28   &	0.30  &	0.22   &	0.27   &	0.23   &	0.24  & +0.25$\pm$0.01   & 4.94\\
$ [V/Fe] $   &	 -0.14  &	-0.11  &	-0.23  &	-0.08 &	-0.10  &	-0.04  &	-0.12  &	-0.09 & -0.11$\pm$0.02  & 4.00\\
$ [CrI/Fe] $  &	 0.02   &	0.05   &	-0.01  &	0.03  &	-0.02  &	0.01   &	-0.05  &	-0.08 & -0.01$\pm$0.01   & 5.63\\
$ [Mn/Fe] $  &	 -0.22  &	-0.20  &	-0.25  &	-0.27 &	-0.15  &	-0.16  &	-0.27  &	-0.26 & -0.22$\pm$0.02   & 5.37\\
$ [Fe/H] $   &	 -1.51  &	-1.53  &	-1.60  &	-1.48 &	-1.49  &	-1.55  &	-1.52  &	-1.55 & -1.53$\pm$0.01   & 7.50\\
$ [Co/Fe] $  &	 0.26   &	0.16   &	0.13   &	0.13  &	0.18   &	0.21   &	0.07   &	0.22  & +0.17$\pm$0.02   & 4.93\\
$ [Ni/Fe] $  &	 -0.08  &	-0.08  &	-0.03  &	-0.07 &	-0.13  &	-0.09  &	-0.11  &	-0.07 & -0.08$\pm$0.01  & 6.26\\
$ [Cu/Fe] $  &	 -0.69 &	-0.36  &	-0.26  &	-0.4 &	-0.52  &	-0.47  &	-0.39  &	-0.54 & -0.45$\pm$0.05   & 4.19\\
$ [Zn/Fe] $  &	 -0.30  &	-0.19  &	-0.16  &	-0.17 &	-0.31  &	-0.26  &	-0.15  &	-0.22 & -0.22$\pm$0.02   & 4.61\\
$ [Y/Fe] $   &	 -0.19  &	-0.15  &	-0.28  &	-0.14 &	-0.21  &	-0.10  &	-0.11  &	-0.11 & -0.16$\pm$0.02   & 2.25\\
$ [Zr/Fe] $  &	 0.19   &	0.21   &	0.23   &	0.43  &	0.17   &	0.27   &	0.16   &	0.31  & +0.25$\pm$0.03   & 2.56\\
$ [Ba/Fe] $  &	 0.12&	0.15  &	0.07  &	0.19 &	0.13  &	0.08  &	0.14  &	0.13 &   +0.13$\pm$0.01   & 2.34\\
$ [La/Fe] $  &	 0.07   &	0.01   &	-0.11  &	-0.01 &	0.02   &	0.14   &	0.09   &	0.09  & +0.04$\pm$0.03  & 1.26\\
$ [Ce/Fe] $  &	 -0.29  &	-0.16  &	-0.24  &	0.06  &	-0.18  &	-0.22  &	-0.21  &	-0.07 & -0.16$\pm$0.04  & 1.53\\
$ [Pr/Fe] $  &	 0.19   &	0.19   &	0.07   &	0.37  &	0.17   &	0.23   &	0.32   &	0.23  & +0.22$\pm$0.03   & 0.71\\
$ [Nd/Fe] $  &	 0.13   &	0.12   &	-0.12  &	0.31  &	0.16   &	0.21   &	0.25   &	0.19  & +0.16$\pm$0.05   & 1.59\\
$ [Sm/Fe] $  &	 0.18   &	0.21   &	0.17   &	0.46  &	0.28   &	0.31   &	0.21   &	0.30  & +0.27$\pm$0.04   & 0.96\\
$ [Eu/Fe] $  &	 0.28   &	0.33   &	0.25   &	0.51  &	0.41   &	0.39   &	0.44   &	0.38  & +0.37$\pm$0.03   & 0.52\\
$ [Dy/Fe] $  &	 -0.15  &	0.2  &   -	  &	0.57  &	-0.04  &	-0.22  &	0.01 &	-0.4 & 0.00$\pm$0.12& 1.10\\
$ [Pb/Fe] $  &    -           & 	 0.03 &	0.26   &	0.31 &	-0.13  &	-0.11  &	-0.09  &	0.21  & +0.07$\pm$0.07   & 1.98\\
\hline

\end{tabular} 
  \begin{tablenotes}
   \item[1] The errors are statistical  errors obtained of the mean.
   \item[2]  [Ti/Fe] is the average  between Ti I and  Ti II.
  \end{tablenotes}
\end{threeparttable}
\end{table*}

\begin{table*}
\caption{Estimated errors on abundances, due to errors on atmospherics parameters and to spectral noise, compared with the observed errors.}
\label {t3}
\centering
\begin{tabular}{ l c  c c  c  c c  c  c c  c }
\hline 
\hline
	ID   &  $\Delta T_{eff} =40 K $ & $ \Delta log(g)=0.14$  & $\Delta v_{t}= 0.02$	& $ \Delta [Fe/H]=0.04 $ & S/N & $\sigma_{tot}$  & $\sigma_{obs}$\\		
	\hline					
	$ \Delta ([C/Fe]) $     &	-0.06  &	-0.03  &	-0.01  &	0.01   &	0.04  &	0.08  &	0.25\\ 	
	$ \Delta ([N/Fe]) $     &	-0.06  &	-0.01  &	-0.05  &	-0.01  &	0.04  &	0.09  &	0.38\\ 	
	$ \Delta ([O/Fe]) $     &	-0.02  &	0.05   &	0.01   &	0.04   &	0.05  &	0.08  &	0.26\\ 	
	$ \Delta ([Na/Fe]) $  &	-0.02  &	-0.01  &	0.00   &	0.01   &	0.04  &	0.04  &	0.35\\ 	
	$ \Delta ([Mg/Fe]) $    &	-0.03  &	0.03   &	0.00   &	0.01   &	0.04  &	0.06  &	0.07\\ 	
	$ \Delta ([Al/Fe]) $    &	-0.01  &	-0.02  &	0.04   &	0.02   &	0.03  &	0.06  &	0.34\\ 	
	$ \Delta ([Si/Fe]) $    &	0.01   &	-0.07  &	0.00   &	0.00   &	0.05  &	0.09  &	0.06\\ 	
	$ \Delta ([Ca/Fe]) $    &	0.00   &	-0.03  &	-0.01  &	-0.01  &	0.04  &	0.05  &	0.04\\ 	
	$ \Delta ([Sc/Fe]) $    &	0.02   &	-0.05  &	-0.01  &	0.01   &	0.05  &	0.07  &	0.10\\ 	
	$ \Delta ([Ti I/Fe]) $  &	0.03   &	-0.02  &	0.00   &	-0.01  &	0.02  &	0.04  &	0.05\\ 	
	$ \Delta ([Ti II/Fe]) $ &	-0.05  &	0.04   &	-0.01  &	-0.01  &	0.09  &	0.11  &	0.06\\ 	
	$ \Delta ([V/Fe]) $     &	-0.01  &	-0.02  &	0.01   &	0.03   &	0.03  &	0.05  &	0.06\\ 	
	$ \Delta ([Cr I/Fe]) $  &	0.03   &	-0.02  &	-0.01  &	-0.01  &	0.03  &	0.05  &	0.04\\ 	
	$ \Delta ([Mn/Fe]) $    &	0.00   &	-0.02  &	0.00   &	-0.01  &	0.04  &	0.05  &	0.05\\ 	
	$ \Delta ([Fe/H]) $     &	0.05   &	0.01   &	0.00   &	0.00   &	0.01  &	0.05  &	0.04\\ 	
	$ \Delta ([Co/Fe]) $    &	0.02   &	0.00   &	-0.01  &	-0.02  &	0.04  &	0.05  &	0.06\\ 	
	$ \Delta ([Ni I/Fe]) $  &	-0.01  &	0.00   &	0.00   &	0.00   &	0.02  &	0.02  &	0.03\\ 	
	$ \Delta ([Cu/Fe]) $    &	-0.04  &	-0.02  &	0.00   &	-0.03  &	0.04  &	0.07  &	0.13\\ 	
	$ \Delta ([Zn/Fe]) $    &	-0.01  &	0.03   &	0.02   &	0.06   &	0.04  &	0.08  &	0.06\\ 	
         $ \Delta ([Y/Fe]) $     &	0.01   &	0.03   &	-0.02  &	0.00   &	0.04  &	0.05  &	0.06\\  	
	$ \Delta ([Zr/Fe]) $    &	-0.05  &	0.06   &	0.01   &	0.01   &	0.03  &	0.08  &	0.09\\ 	
	$ \Delta ([Ba/Fe]) $    &	-0.05  &	0.06   &	0.00   &	0.01   &	0.04  &	0.09  &	0.04\\ 	
	$ \Delta ([La/Fe]) $    &	-0.03  &	0.04   &	0.00   &	0.01   &	0.04  &	0.06  &	0.08\\ 	
	$ \Delta ([Ce/Fe]) $    &	-0.01  &	0.03   &	0.00   &	0.02   &	0.05  &	0.06  &	0.11\\ 	
	$ \Delta ([Pr/Fe]) $    &	-0.01  &	0.02   &	-0.01  &	0.00   &	0.04  &	0.05  &	0.09\\ 	
	$ \Delta ([Nd/Fe]) $    &	-0.06  &	0.04   &	-0.03  &	0.00   &	0.04  &	0.09  &	0.13\\ 	
	$ \Delta ([Sm/Fe]) $    &	0.01   &	0.01   &	0.00   &	0.02   &	0.04  &	0.05  &	0.10\\ 	
	$ \Delta ([Eu/Fe]) $    &	-0.06  &	0.06   &	0.00   &	0.01   &	0.04  &	0.09  &	0.09\\ 	
	$ \Delta ([Dy/Fe]) $    &	-0.03  &	0.05   &	0.01   &	0.02   &	0.04  &	0.07  &	0.32\\ 	
	$ \Delta ([Pb/Fe]) $    &	-0.19  &	-0.17  &	-0.18  &	-0.10  &	0.04  &	0.33  &	0.19\\	
	
\hline

\end{tabular}
\end{table*}

%-----------------------------------------------------------------------------------------------------------------------------------------------------------------------
%-------------------------------------------------------------------------------------RESULTS--------------------------------------------------------------------
%-----------------------------------------------------------------------------------------------------------------------------------------------------------------------
\section{Results}
\label{results}

In the following section, we discuss in detail our results and compare them  with the literature.

%-----------------------------------------------------------------------------------------------------------------------------------------------------------------------
%---------------------------------------------------------------------IRON AND IRON-PEAK ELEMENTS--------------------------------------------------
%-----------------------------------------------------------------------------------------------------------------------------------------------------------------------

\subsection{Iron-peak elements}
\label{iron}

We found a mean  [Fe/H] value for NGC3201 of:
\begin {center}
\vspace{0.3cm}
[Fe/H]=$-1.53\pm0.01$ dex
\vspace{0.3cm}
\end {center}

\noindent The observed scatter is entirely consistent with that expected solely from errors and thus we find no
evidence for any intrinsic Fe abundance spread, except for star 3 of our sample (analyzed in more  detail below).
\citet{gonzalez} found a mean [Fe/H]=$-1.42\pm0.12$ from their high resolution analysis of 
eighteen stars. Some of their stars showed Fe abundances significantly different from the mean,
but they associated this behavior as likely due to a systematic error in the abundance determination and not a real metallicity spread. Although \citet{covey}  cannot 
confirm a spread of iron  in  NGC 3201 greater than 0.3 dex.
\citet{carretta2009c} used GIRAFFE at the VLT to obtain
high resolution spectra for 149 giants. They found a mean  [Fe/H]=$-1.50\pm0.07$ with
   $\sigma=0.049$. In addition, their sample of 13 even higher resolution UVES stars yielded a mean [Fe/H]=$-1.51\pm0.08$ with  $\sigma=0.07$.  Their combined sample is impressively  large and homogeneously analyzed
and they did not find any spread.  The results from Carretta are in excellent agreement with ours, both in mean [Fe/H]
as well as the lack of an Fe abundance spread, giving us added confidence in our results.

On the other hand, \citet {simmerer} recently  found an intrinsic spread of iron  of 0.4 dex in a sample of 24 RGB stars.
However,  when  we analyze the photometry  of  the metal poor stars  from \citet {simmerer}, we see  that 5 of the 6 stars with  [Fe/H]$\le$-1.59 (filled cyan triangles in Figure 3) are most likely AGB and not RGB stars, as shown in  Figure  3.   Stars with [Fe/H]$\geq$-1.56  from \citet {simmerer} are essentially all RGB stars.\\

It is noteworthy that when comparing the 149   stars from \citet{carretta2009c}   with the data from \citet {cote}, which cover the full color range of the RGB, we note that  Carretta covered RGB fairly well, giving  reliability to  our data.\\

 We further investigate any potential met spread in our sample by comparing stars \#3 and \#4 of our sample. Both stars have a similar temperature and  the most significant difference in [Fe/H], 0.12 dex. Figure 4 shows the star more metal-poor in black  and  the more metal-rich in red, both spectral lines are Fe. There is a difference  in the depth of the lines, which could  be attributed to an intrinsic spread, as found  by \citet {simmerer}. Also in the Figure 4 we compare a pair of stars with similar Fe abundance and temperature, these do not show any  difference in the depth of these lines.
On the other hand star \#3 is the only star  of our sample that  shows this difference, therefore our result is not conclusive. We cannot rule out a variation on the order of 0.12 dex.\\
Although there is evidence for an intrinsic spread of iron in NGC3201, mainly found by \citet {simmerer}, this needs to be studied in more detail. Because
so far, only the more massive GCs have  shown a larger  spread in iron \citep{carretta2009c}. However, NGC3201 is among the less massive halo clusters.

The chemical abundances for the Iron-peak elements Sc, V, Cr, Mn, Co, Ni, Cu and Zn listed in Table 2  are solar scaled or slightly 
underabundant  in the cases of Mn, Zn  and especially Cu. 
Figure 5 shows each of our stars in each of these elements compared with results for the halo field stars, disk stars, other GCs, LMC stars,  dSphs, etc. In general, we found  that NGC 3201 stars have abundances of these elements which agree with those of other GCs and halo field
stars of similar metallicity. All elements, with the possible exception of V, are consistent with no intrinsic dispersion.

Our only strongly non-solar Fe-peak element is Cu, with a value of [Cu/Fe] = $-0.45 \pm0.05$. 
The origin of Cu has been discussed by several authors \citep{matteucci,raiteri,sneden1991}. It can be produced by different nucleosynthetic processes and it is, therefore, difficult to give a definitive explanation of the origin of this element in GCs.

\citet{sneden1991} suggests that the formation of Cu took place as a weak component of the s-process in massive stars.
\citet{matteucci}  and \citet{raiteri}  suggest that a strong contribution of Cu comes  from SNeIa. Therefore, Cu could be a good indicator of the pollution due  to the  SNeIa to the cluster.

When analyzing the possible origin of Cu in NGC3201, we  found that there is no significant correlation between Cu and weak-s component elements (Y, Zr) nor between Cu and main-s components (Ba). Therefore,  we can rule out a significant contribution of these processes to the nucleosynthesis of Cu. On the other hand, given the low abundance of Cu and Zn (the next heavier element), the low spread in both  and  the overabundance of the alpha elements, we suggest a low contribution of SNeIa to NGC3201.

%-----------------------------------------------------------------------------------------------------------------------------------------------------------------------
%------------------------------------------------------------------------ALPHA -ELEMENTS------------------------------------------------------------------
%-----------------------------------------------------------------------------------------------------------------------------------------------------------------------

\subsection{$\alpha$ elements}
\label{alpha}
All the $\alpha$ elements (O, Mg, Si, Ca  and Ti) listed in Table 2  are overabundant.
This is typical of almost every GC as well as  similarly metal-poor  halo field stars  in the Galaxy, as shown in Figure 13. This is our first strong chemical suggestion that NGC 3201 may NOT have an extraGalactic origin, or that at least its chemical evolution has been very similar to that of other GCs.

If we leave out O (due to its involvement in the Na:O anticorrelation addressed below) and use only Mg, Si, Ca and Ti, we derive a mean $\alpha$ element abundance for NGC3201 of

\begin {center}
\vspace{0.2cm}
[$\alpha$/Fe]=$0.30\pm0.06$
\vspace{0.3cm}
\end {center}

There is good agreement between $ \sigma_{tot}$ and $\sigma_{obs}$, and we conclude that there is no evidence of any internal spread in the $\alpha$ elements. 
Our mean value is in good agreement with \citet{gonzalez}, who found  [$\alpha$/Fe] = $0.36\pm0.09$, and \citet{carretta2009a,carretta2010a}, who found a mean value of [(Mg+Ca+Si)/Fe] = $0.31\pm0.02 $.

Virtually all GGCs show a similar overabundance of the  alpha elements,  as shown in Figures 12 and 13. This behavior is due to the contribution of high-mass stars that end their lives as SNeII, which are very  efficient  in  enriching the interstellar medium with alpha elements \citep{tinsley,sneden2004}.

 From Figure 13 we notice that GGCs (in blue) for Mg, Si, Ti and specially Ca  show a very strong trend, in each case NGC3201 follows the same trend.\\

If we consider the following points  about the  $\alpha$ elements, it  is possible to understand their  behavior in NGC3201 :
\begin{itemize}
  \item Proton-capture nucleosynthesis  can affect the abundance of  Mg, but this element shows no significant spread, although Al does.
   \item Ti can  be affected  by other nucleosynthetic processes, but this element also shows no significant spread.
   \item The main source of $\alpha$-elements are  SNeII, and   the pure alpha-elements are Ca and Si.
    \item    The mean abundance we find for pure alpha-elements is   [(Ca+Si)/Fe]=0.29, which is consistent with the value of  all alpha elements (Mg, Si, Ca and Ti).  
   \end{itemize}
 The enhanced [$\alpha$/Fe] ratio we find in NGC 3201, closely mimicking values found in other Galactic halo tracers, indicates that it also experienced rapid chemical evolution dominated by SNeII. This strongly suggests that its chemical enrichment was very similar if not identical to that experienced in
these other halo objects and was not like that of dSphs, which experienced much slower chemical
evolution, leading to low and even subsolar [$\alpha$/Fe]  ratios at low metallicity \citep[e.g.][]{geisler}.

%-----------------------------------------------------------------------------------------------------------------------------------------------------------------------
%-------------------------------------------------------------------NA-O ANTICORRELATION---------------------------------------------------------------
%-----------------------------------------------------------------------------------------------------------------------------------------------------------------------

\begin{figure}
\centering
    \includegraphics[width=3.2in,height=3.2in]{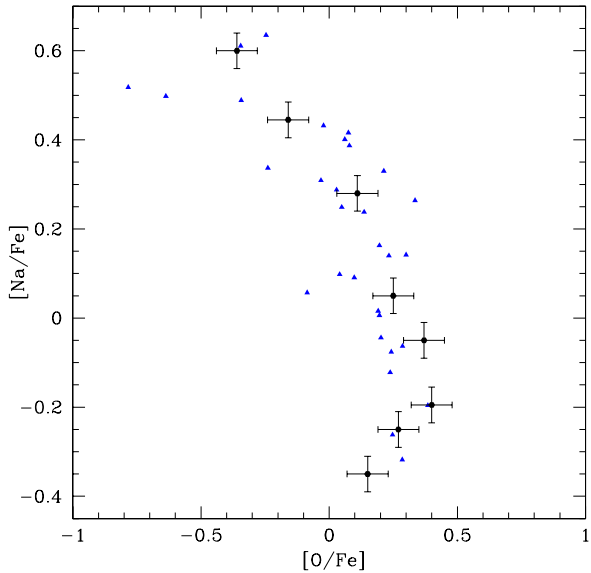}
  \caption{ Na-O anticorrelation for NGC3201. Filled black circles with error bars are our data. Filled blue triangles are  data from \citet{carretta2009b}.}
  \label{f2}
 \end{figure}

\subsection{Na-O anticorrelation}
\label{Na-O}
The anticorrelation between Na and O has been very well established and studied in many GCs, including NGC 3201.
Both \citet{gonzalez} and \citet{carretta2009b} have in fact already established this 
anticorrelation in our cluster. We completely corroborate their findings.
In Figure 6 our data appear together with those of Carretta (2009b). Very good agreement occurs between them.

 \citet{carretta2009a}  found spreads in O and Na very similar to ours: $ \sigma_ {obs Carretta} [O / Fe]= 0.28 $ and $ \sigma_ {obs Carretta} [Na / Fe]= 0.31 $.
The dispersion in [O/Fe] and especially  [Na/Fe] is one of the highest among the 19 GCs studied by \citet{carretta2009a}, 
 comparable to those in  NGC1904, NGC2808 and NGC6752. 
 
\citet{carretta2007} found a strong relation between Na-O extension and the extension of the HBs, in agreement  with   \citet{dantona}  prediction. In the case of NGC3201 it shows an extended Na-O anticorrelation with a well populate and extended HB.

  Our data do suggest a curious trend for the most Na-poor stars to actually show a small Na:O correlation, but this is based on only three stars and is very preliminary.

 Our sample suggests stars of the first generation, generally believed to be those lowest in [Na/Fe] and highest in [O/Fe], have abundances for both of these elements relatively low compared with the other GCs from  \citet{carretta2009a}. This could indicate that NGC 3201 formed in an environment with exceptionally low
Na and O for its metallicity compared to other GCs, a hint in favor of an extraGalactic origin, but more data are needed to clarify this issue, as we have only one star (star \#3) which is  very Na poor. An example of the spectral fit to Na  in star \#3 is given in Figure 7.

\begin{figure}
\centering
    \includegraphics[width=3.1in,height=3.8in]{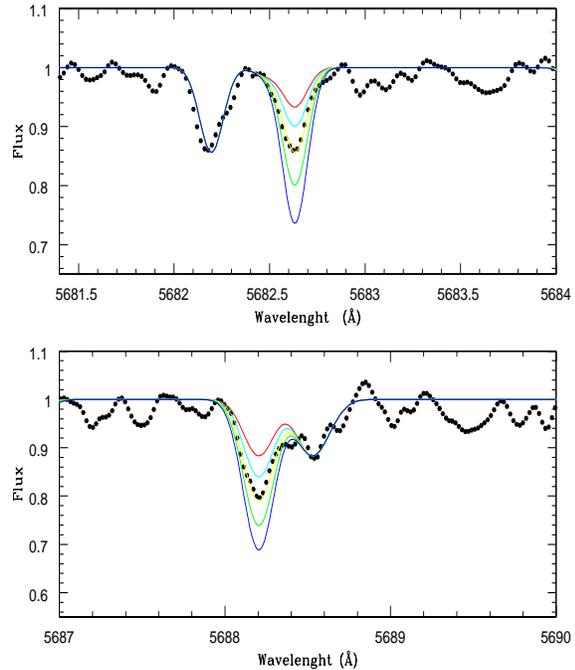}
  \caption{Spectrum synthesis fits for two lines of Na in the star \#3 from NGC3201. The  solid color lines show the synthesized spectra corresponding to different abundances of Na ([Na/Fe]=0.12, -0.08, -0.32, -0.47, -0.67).}
  \label{f2}
 \end{figure}

\begin{figure}
\centering
    \includegraphics[width=3.1in,height=3.8in]{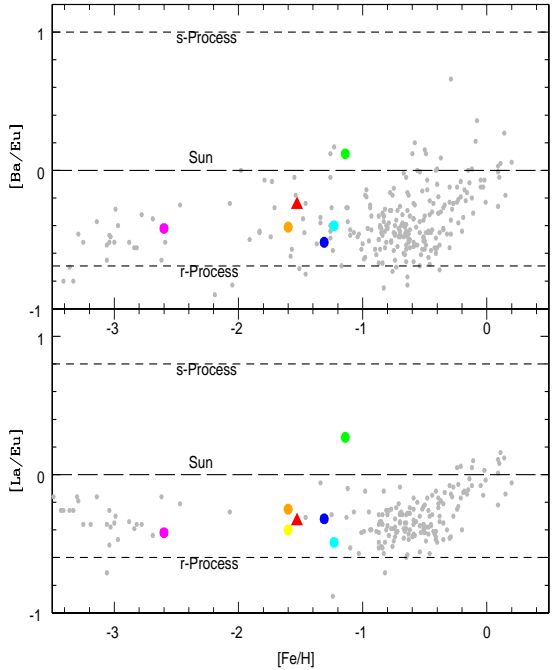}
  \caption{ [Ba/EU] and  [La/Eu] vs [Fe/H]. The filled red  triangle is our mean for NGC 3201. Filled gray circles are halo and disk stars from  \citet{fulbright} and \citet{francois}, The green circle is M4 \citep{yong2008}; the magenta circle is M15 \citep{sobeck}; the blue circle is M5 \citep{yong2008}; the orange circle is NGC6752 and  the yellow circle is M13 from \citet{yong2006} and the cyan circle is NGC1851 from \citet{villanova2010}.}
  \label{f2}
 \end{figure}

\begin{figure}
\centering
    \includegraphics[width=3.2in,height=3.2in]{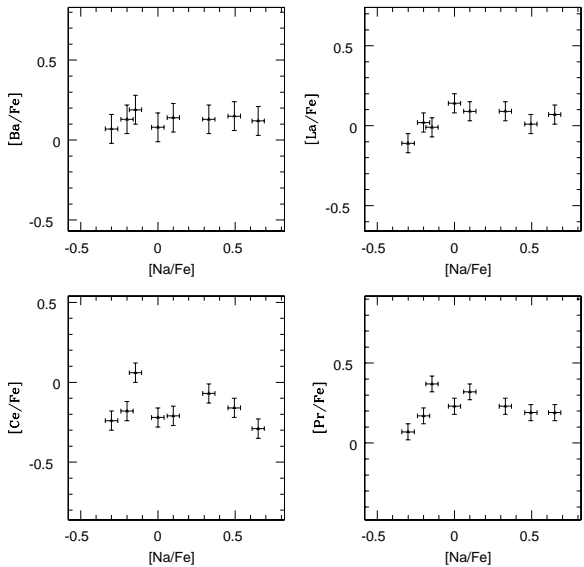}
  \caption{Upper panels: [Ba/Fe] and [La/Fe] as a function of [Na/Fe] for our data. Lower panels: [Ce/Fe] and [Pr/Fe] as a function of [Na/Fe] for  our data.}
  \label{f2}
 \end{figure}

%-----------------------------------------------------------------------------------------------------------------------------------------------------------------------
%-------------------------------------------------------------Mg-Al--------------------------------------------------------------------
%-----------------------------------------------------------------------------------------------------------------------------------------------------------------------

\subsection{Mg-Al anti-correlation}
\label{Mg-Al}

A Mg-Al anticorrelation is also found in some GCs \citep{carretta2009a}. However, unlike the O-Na anticorrelation, a Mg-Al anticorrelation  is more difficult to reproduce in simulations and it remains  a puzzle  to explain the possible sources that would produce this anticorrelation \citep{denissenkov,yong2003}.\\
NGC3201 exhibits a curious behavior in this regard. Our sample has  a significant spread in Al ($\sigma_{obs}=0.34\pm0.12$) but no significant spread in Mg ($\sigma_{obs}=0.08\pm0.03$ ), Figure 10 shows our data  together with data from \citet{carretta2009a}  and \citet{gonzalez}. Although the observed range in Mg abundance is only slightly larger than that expected from observational errors, one could imagine a behavior in this diagram similar to that hinted at in the Na:O diagram, in which the stars
with the smallest [Al/Fe] abundances exhibit a suggestion of a correlation, followed by a possible 
anticorrelation for the most [Al/Fe] enhanced stars.

 The possible "correlation" seen in  the first stars in NGC3201 is similar to the behavior displayed by disk and halo field  stars. A shift to a weak anticorrelation in NGC3201 appears to occur  approximately at the  maximum value for Mg and Al reached by stars in the halo and the disk, although the NGC 3201 sample are generally displaced to smaller [Al/Fe] vales at the same [Mg/Fe]. Although the explanation for this behavior is not clear, if supports the possibility of a Galactic origin for NGC3201.

   \begin{figure}
\centering
    \includegraphics[width=3.2in,height=3.2in]{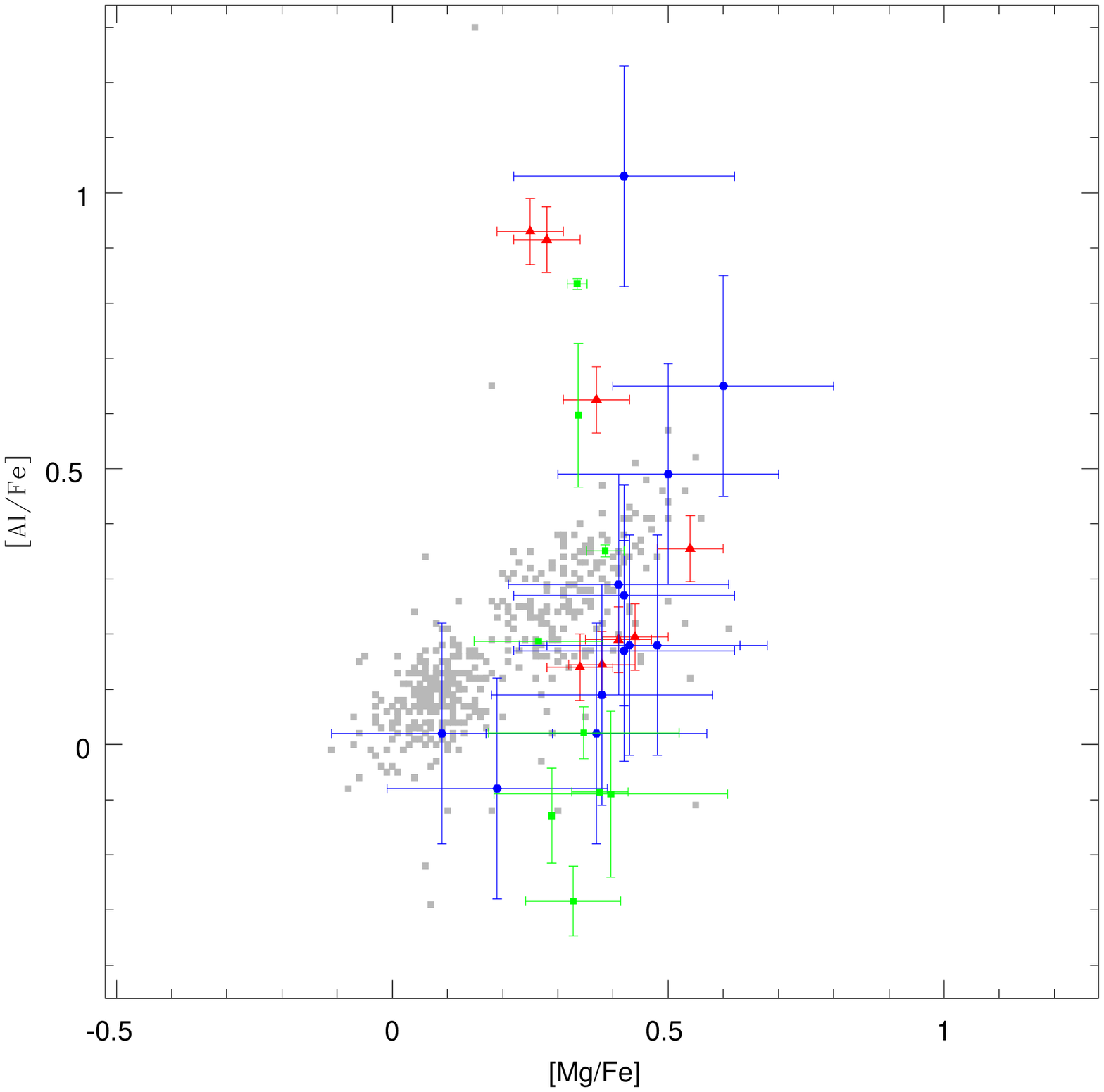}
  \caption{[Mg/Fe] vs [Al/Fe]. Filled red triangles are our data for NGC3201,  filled blue circles are data from \citet{gonzalez} for NGC3201, filled green squares are data from \citet{carretta2009a} for NGC3201; Filled gray squares are halo and disk field stars from \citep{fulbright,reddy2003,cayrel,barklem,reddy2006}.}
  \label{f2}
 \end{figure}

%-----------------------------------------------------------------------------------------------------------------------------------------------------------------------
%-------------------------------------------------------------CNO AND AGE CORRECTION----------------------------------------------------------------
%-----------------------------------------------------------------------------------------------------------------------------------------------------------------------

\subsection{C+N+O and Age}
\label{CNO}

We measured the total (C+N+O) content in NGC3201, obtaining a mean of [(C+N+O)/Fe]= 0.18 dex, with an insignificant spread.\\

The total (C+N+O) content has now been measured in several Galactic GCs:
M4 stars have  [(C+N+O)/Fe]=+0.29 dex \citep{villanova2011},  
NGC1851 giants show no significant  difference in (C+N+O) between the Ba-rich ( [(C+N+O)/Fe]=+0.20 dex)  and Ba-poor ([(C+N+O)/Fe]=+0.23 dex)
populations \citep{villanova2010}  but  \citet{yong2009}   found a spread of  $\sim0.6$ dex in NGC1851.
NGC6397, NGC6752  and 47 Tuc also were measured by \citet{carretta2004} and their stars  show a constant (C+N+O) content.  
NGC3201 stars also display  a constant (C+N+O) content. This behavior, apparently, is a trait of the  mono-metallic GCs. 
On the other hand, the GCs non-monometallic show a variation in the content of (C+N+O), e.g. $\omega$ Cen shows a correlation between the CNO and their iron content abundance, increasing the [(C+N+O)]/Fe] by $\sim0.5$ dex between metallicity range from [Fe / H] $\sim-2.0$ to [Fe / H] $\sim-0.9$ \citep{marino2012}. M22 another non-monmetallic  cluster shows a  similar correlation that  $\omega$ Cen \citep{marino2011b}.

The age derived from isochrone fits depends sensitively on (C+N+O) \citep{cassisi}.  \citet{marino2012} found the following relation:

\vspace{0.3cm}
$\delta$ Age/ $\delta$ [C+N+O] $\sim$ -3.3 Gyr/dex
\vspace{0.3cm}

\citet{dotter}   derive an age=$12.0\pm0.75$ Gyr from a deep HST CMD of NGC 3201 assuming  an $\alpha$-enhanced composition.
  This age then should be corrected for the appropriate (C+N+O) abundance we have derived.
According to the equation above, the corrected \hspace {0.1 cm}age of \hspace {0.2 cm}NGC3201 is  -3.3 X 0.18 $\sim$ -0.6 Gyrs. This brings its age
 to $\sim$ 11.4 Gyrs.  The values used by \citet{dotter}  for [Fe/H] and [$\alpha$/Fe]   were only slightly smaller than our values and have a much smaller impact on the derived age.

%-----------------------------------------------------------------------------------------------------------------------------------------------------------------------
%---------------------------------------------------------------------HEAVY ELEMENTS----------------------------------------------------------------------
%-----------------------------------------------------------------------------------------------------------------------------------------------------------------------
\subsection{Heavy elements}
\label{Heavy elements}
We measured a number of heavy elements  (Y, Zr, Ba, La, Ce, Pr, Nd, Sm, Eu, Dy and Pb). These elements are mainly produced by two
 processes: slow\hspace {0.1cm} neutron capture \hspace {0.1cm} reactions (s-process), in which the neutron capture time is much longer than the beta decay lifetime  and  rapid neutron capture reactions (r-process), where neutron capture times are much shorter than the beta-decay lifetime.

The study of these elements provides the opportunity to better understand these nucleosynthetic  
processes in GCs, The s-process occurs primarily in AGB stars, producing mainly light-s  elements such as Sr, Y, Zr, and heavy-s elements 
 like Ba, whereas massive main sequence stars produce only light s-elements. The r-process 
 occurs mainly in SNeII explosions, producing in  particular Eu, along with significant $\alpha$ and iron-peak  element enhancements. Thus, studying the detailed heavy element abundances allows us to determine the relative contribution of these processes to the cluster's chemical evolution, in particular during the formation of the second generation, and possibly shedding significant
light on this important epoch. The analysis of these elements allows us to investigate the nature of  the first generation stars which polluted subsequent generations.
 
 The [Ba/Eu] and [La/Eu]  ratios are  very sensitive to the relative importance of the   
 s-process vs. r-process.   For this purpose, in Figure 8 we plot the abundances of  [Ba/Eu] and [La/Eu] vs [Fe/H]. In both cases, we find that these neutron capture elements show a nucleosynthetic history more dominated \hspace{0.1mm} by the r-process than the s-process. This behavior is also seen in all of the
 other samples displayed except for M4 and a small sample of disk stars.
 
The Ba content is constant, with no intrinsic spread,  suggesting that AGB stars cannot be a  possible polluter, as we should then see the difference between the first and second generations reflected in a spread.
 This is corroborated by analyzing other heavy s-process elements including La, Ce and Pr (Figure 9), which all show a negligible intrinsic spread. 
The light s-elements measured by us (Y, Zr) also show no  significant spread.\\

Star \#3 has other peculiar features, besides being the most metal poor of our sample (see section 4.1). It also  has the lowest values  for Y, Ba, La, Pr, ND, Sm, Y and Eu. In some cases their  values are very extreme, in particular Y, La and Nd. Some GCs like M22 \citep{marino2011b} and NGC1851 \citep{carretta2011} have shown similar features, where more iron rich stars are also  more neutron-capture rich. This potentially supports the intrinsic spread in Fe in NGC3201 found by \citet {simmerer}. On the other hand, in the case of our study we found this feature in only one star of  our sample, and although the result  is not conclusive it is important to bear in mind.\\

Another point  that is striking is the value of  sigma estimated and sigma observed for Dy. The difference is  almost  3 sigma, which  could be interpreted as a spread of Dy in NCG3201. However, no other neutron capture elements show this spread. Moreover the  value in   star \#4  is unusually high among our sample. Another factor  that could  affect the measurement of Dy is that the abundance  was determined using only one line (5169.688 $\AA$)  and this  line  is very weak. A larger sample of stars with better determined values of Dy would be useful.

\begin{figure*}
\centering
    \includegraphics[width=5.2in,height=5.2in]{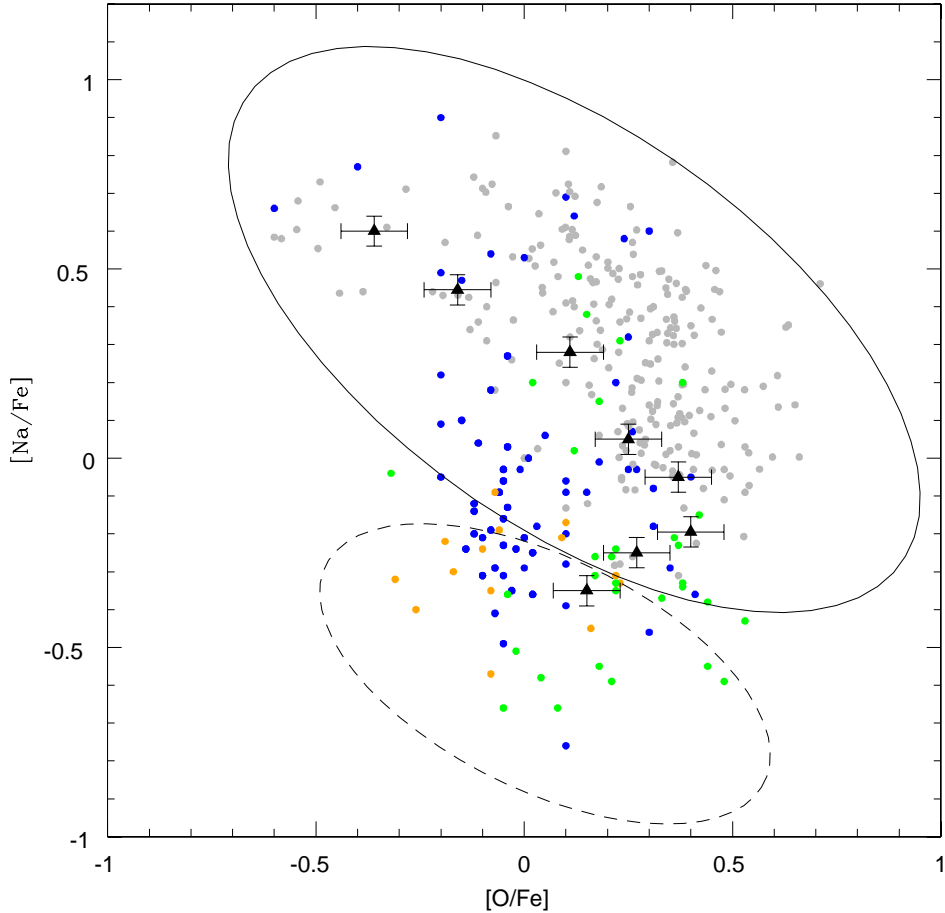}
  \caption{ Na-O anticorrelation. Filled black triangles with error  bars are our data from NGC3201. Filled gray squares are 19 GCs from \citet{carretta2009a,carretta2009b}. Blue filled circles are objects from LMC \citep{johnson,pompeia,mucciarelli2008}. Filled orange circles are sagittarius dwarf  galaxy \citep{sbordone}. Filled green circle are Draco, Sextans and Ursa minor dwarf galaxies \citep{shetrone}. }
  \label{f2}
 \end{figure*}

%-----------------------------------------------------------------------------------------------------------------------------------------------------------------------
%------------------------------------------------------------ANALYSIS ORIGIN NGC3201------------------------------------------------------------------
%-----------------------------------------------------------------------------------------------------------------------------------------------------------------------

\section{Analysis of the origin and chemical evolution of NGC3201}
\label{Origin}

The origin of NGC 3201 is of great interest, not only for its own sake as an intriguing GC but also for helping to understand the origin
of the Galactic GC system in general. In particular, we wish to address the Galactic vs. extraGalactic origin of this exotic GC. Additionally, we would like to compare its chemical evolution to that of other GGCs and help constrain scenarios for the generation of MPs. 

Kinematically, it is well known that this cluster is an extreme object.
It has\hspace {0.3 cm} a very \hspace {0.2 cm}large \hspace {0.2 cm}radial velocity, $495$ $km$ $s^{-1}$, the largest among GGCs. In addition, it has  a retrograde orbit. Both of these factors strongly suggest an extraGalactic origin.
\citet{rodgers} and  \citet{van}  included it in a list of objects that possibly formed during the merging of the Galaxy with  other objects. 
Allen et al. (2008) studied  six globular clusters to determine their galactic orbits (NGC2808, NGC3201, NGC4372, NGC4833, NGC5927, NGC5986) and   found that NGC3201 is the  least bound to the bar. They  showed that its orbit is always outside the bar region using an  axisymmetric and the barred potentials representative of our galaxy.

Our analysis, as well as that of\hspace {0.1cm} both Gonzalez and Carretta before us, has been done  from a chemical point of view, adding a distinct dimension to our understanding of the cluster and its possible origin. In particular, we can compare its detailed chemical behavior to that of other  objects with a variety of suspected origins. We first replot the Na-O anticorrelation (Figure  11)  where we now compare our data (points  with error bars) with that of 19 other Galactic GCs  \citep{carretta2009a,carretta2009b}. In addition, we include LMC objects \citep{johnson,pompeia,mucciarelli2008}, Sagittarius dwarf galaxy field and cluster star \citep{sbordone}, Draco, Sextans and Ursa Minor dwarf galaxy field star \citep{shetrone}. 

We can clearly see a region that is only occupied   by  extraGalactic objects (enclosed by a dashed ellipse) and another region occupied  mainly  by GGCs (enclosed by a continuous  ellipse). 
The bulk of NGC 3201 giants lie within the Galactic region. However, the trend for the lowest Na stars
in NGC 3201 to show a decreased O abundance compared to somewhat higher Na stars is rather unique amongst GGCs, and indeed places these stars, which would generally be considered the likely first generation stars, at the border of the extraGalactic region. Carretta  also found  stars very poor in Na (Figure 6).  These peculiar values  among GGCs opens the possibility that the cluster has an 
extraGalactic origin and subsequently underwent "normal" chemical evolution like typical Galactic GCs,
although this is only a rough guess at this point.
Another interesting fact is that NGC 3201 stars  display a very large intrinsic spread in both Na and O.

Additional evidence for its origin and evolution comes from the [$\alpha$/Fe] vs [Fe/H] diagram (Figure 12). Here we include  the same objects as in Figure 11,  as well as field halo
 and disk stars \citep{fulbright,reddy2003,reddy2006}. [$\alpha$/Fe] was defined as the mean abundance of Mg, Si, Ca and Ti. 
 For LMC's GCs from \citet{mucciarelli2009}, we used only Mg like a representative value to alpha. For red giant branch stars  in  Carina dsph galaxy from \citet{lemasle}  we used  the mean abundance of Mg, Ca and Ti as value to alpha.
  In the case of GGCs from  \citet{carretta2009a},  we added a typical value of Ti for  Galactic GCs  to obtain the value of [$\alpha$/Fe].
  
 In the Figure 12  we again find that Galactic and  extraGalactic objects generally occupy different regimes, with some overlap (dashed line in Figure 12  is the galactic trend and solid line  is extraGalactic trend). 
 NGC3201's members fall in the GGC regime  but nearer the  overlap between galactic and extraGalactic objects than typical GCs.The enhanced [$\alpha$/Fe] of NGC 3201 at its metallicity, along with that of other GCs at other metallicities, graphically illustrates 
their traditional fast chemical evolution, with both Fe and  $\alpha$ elements coming solely from SNeII.
Dwarf spheroidal galaxies, on the other hand, had longer star formation time scales, allowing the  products of SNeIa to eventually deplete the enhancement of  the $\alpha$ elements while continuing to produce additional Fe \citep[e.g.][]{geisler}. The enhanced $\alpha$ abundances, like those in other  GCs, argue against a special origin for NGC 3201, although it does appear to be slightly underabundant
in the alpha elements than typical GCs at its metallicity. From Figure 13, the underabundance principally
stems from Mg and Si, while Ca and Ti appear to be perfectly normal.

From a detailed study of n-capture elements,  we found that the abundances of  [La/Eu] and [Ba/Eu] in NGC 3201 are perfectly normal for a halo GC of its [Fe/H]. The preponderance of the chemical
evidence then clearly suggests that NGC 3201 is a normal halo GC, with the only hints of anything strange coming from the behavior of the most Na-poor stars and the behavior from Mg-Al. We conclude that there is no strong chemical reason to support an origin for NGC 3201 that is any different from the bulk of its Galactic counterparts. 

On the other hand, it is impossible to ignore the peculiar  kinematics  of NGC3201. One possible scenario is that it is indeed of extraGalactic origin but  was captured very early in its chemical evolution by the MW and subsequently evolved as a typical GGC displaying the typical chemical patterns. Perhaps the low original O and Na abundances  
are a reminder of its origin. Clearly, although we have shed light on its intriguing nature, there is much
left to be learned.

\begin{figure}
\centering
    \includegraphics[width=3.2in,height=3.2in]{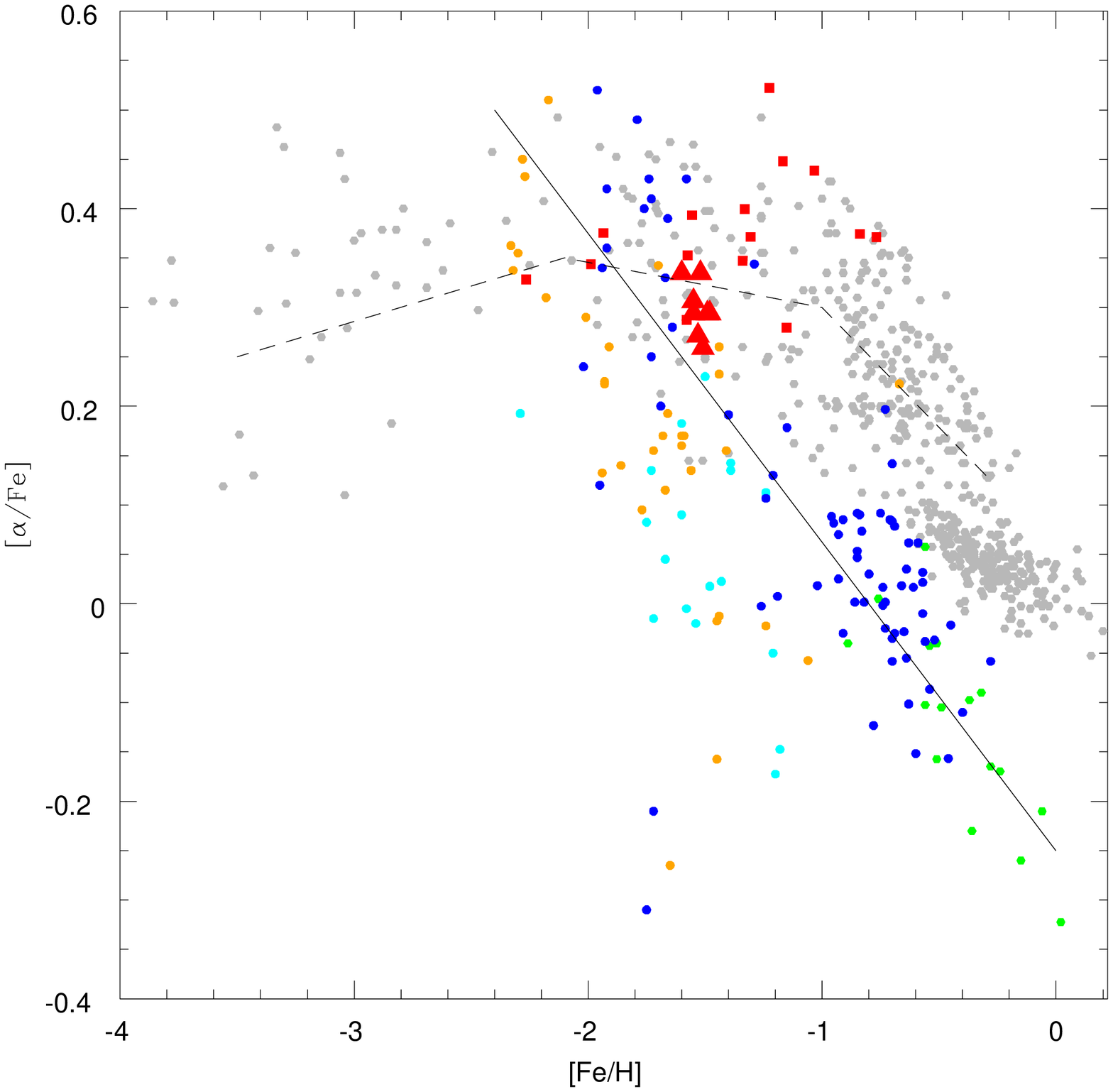}
  \caption{ [$\alpha$/Fe] vs [Fe/H]. Filled red triangles are  our data from NGC3201. Filled gray circle are halo and disk star \citep{fulbright,cayrel,reddy2003,reddy2006}. Filled blue circles are LMC object \citep{johnson,pompeia}. Filled red squares  are galactic GCs \citep{carretta2009a}. Filled orange circles are Draco, Sextans and Ursa minor dwarf galaxy \citep{shetrone}.  Filled green circles are Sagittarius dwarf galaxy \citep{sbordone}. Filled cyan circles  are red giant branch stars  in  Carina dsph galaxy \citep{lemasle}.}
  \label{f2}
 \end{figure}

 \begin{figure}
\centering
   \includegraphics[width=3.1in,height=3.8in]{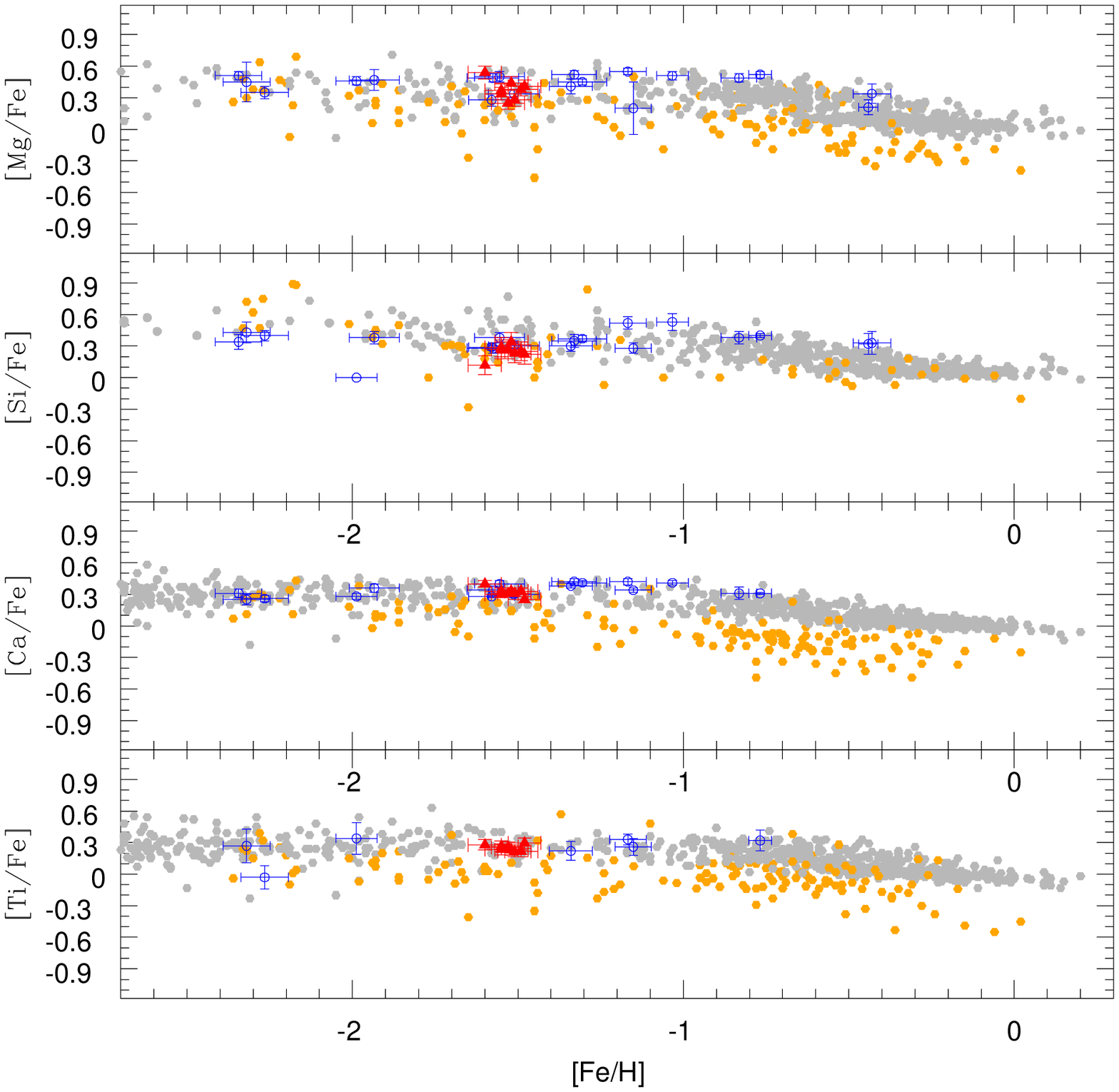}
  \caption{[Mg/Fe],[Si/Fe],[Ca/Fe],[Ti/Fe] vs [Fe/H]. Filled red triangles are giants from NGC3201 from the present study, empty blue circles are GCs \citep{ivans,lee,carretta2006,carretta2009a,villanova2011,koch}. Filled gray circles are halo and disk stars \citep{fulbright,reddy2003,cayrel,barklem,reddy2006}. Filled orange circles are extraGalactic objects Draco, Sextans, Ursa minor and Sagittarius dwarf galaxy \citep{shetrone,sbordone}.}
  \label{f2}
 \end{figure}
 
%-----------------------------------------------------------------------------------------------------------------------------------------------------------------------
%----------------------------------------------------------------------------SUMMARY AND CONCLUSIONS--------------------------------------------
%-----------------------------------------------------------------------------------------------------------------------------------------------------------------------

\section{Summary and Conclusions}
\label{conclusions}
In this paper we present detailed chemical abundances of 29 elements in 8 giant members  of NGC3201 using high resolution, high S/N
 spectroscopy. We measure 4 elements using EW and 25 elements using spectral synthesis and 
 perform an accurate error analysis.\\
We find the following main results: \ \

\begin{itemize}
 \item  NGC 3201 has a mean   [Fe/H] =-1.53$\pm 0.01$ with a $\sigma_{obs}$= 0.04 dex, in good agreement with previous determinations. However, our most metal-poor star may indeed be slightly more metal-poor than the rest. We cannot rule out an intrinsic met spread of about 0.12 dex.
 \item  We confirm the Na-O anticorrelation. Both of these elements show a very large intrinsic spread  compared to the 19 GCs studied by  \citet {carretta2009a,carretta2009b}. We find the most Na-poor stars to also have
 relatively depleted O abundances, hinting that NGC3201 was born in an environment poor in Na.
 \item Intrinsic spreads are also seen in the light elements C, N, and Al but NOT Mg. 
 \item The Fe-peak elements generally show good agreement with other GCs and halo field stars, with no dispersion.

\item  A low Cu abundance together with a normal enhancement of  the alpha elements likely indicates only a small contribution of SNeIa to the pollution in the proto GC.

 \item The alpha elements show the typical overabundance observed in other GCs: [$\alpha$/Fe] =0.30$\pm 0.06$, indicating similar fast star formation time scales and arguing against an extraGalactic
 origin for the cluster, although the mean alpha abundance is slightly lower than that of similar metallicity
 GCs.
 \item The analysis of n-capture elements confirms a dominant contribution from the r-process.
\item The heavy s-process elements like Ba show no significant spread, indicating AGB stars were not
the likely polluters of the second generation.
\item The total (C+N+O) content is constant, similar to most of the monometallic GCs. We derive a correction of some -0.6 Gyr to the age of NGC 3201 derived assuming an $\alpha$-enhanced composition,  yielding $\sim$11.4 Gyr as the corrected value.

\item   Despite its extreme kinematics,  NGC3201 shows no large  chemical differences with respect to other GCs, suggesting its origin is similar to those of other GCs. If it is an extraGalactic object, its chemical evolution was similar to that of other GCs.

 \end{itemize}

%-----------------------------------------------------------------------------------------------------------------------------------------------------------------------
%--------------------------------------------------------------------acknowledgements------------------------------------------------------------------------
%-----------------------------------------------------------------------------------------------------------------------------------------------------------------------

\section*{acknowledgements}

We gratefully acknowledge support from the Chilean 
BASAL   Centro de Excelencia en Astrof\'{i}sica
y Tecnolog\'{i}as Afines (CATA) grant PFB-06/2007.\\
C. Mu\~{n}oz is supported  by CONICYT(Chile)  through
 Programa Nacional de Becas de Magister 2012 (22121324).
We would also like to thank the referee for his  valuable comments.
%-----------------------------------------------------------------------------------------------------------------------------------------------------------------------
%-----------------------------------------------------------------------------BIBLIOGRAPHY--------------------------------------------------------------------
%-----------------------------------------------------------------------------------------------------------------------------------------------------------------------

\bibliographystyle{mn2e}

\label{lastpage}

\end{document}